\newcommand{\be}{\begin{equation}}
\newcommand{\ee}{\end{equation}}
\newcommand{\bea}{\begin{eqnarray}}
\newcommand{\eea}{\end{eqnarray}}
\newcommand{\nn}{\nonumber \\}
\newcommand{\p}[1]{(\ref{#1})}
\documentstyle[12pt]{article}
=0.5cm \oddsidemargin =0.5cm 
\def\a{\alpha}\def\b{\beta}\def\l{\lambda}\def\g{\gamma}\def\z{\zeta}
\def\d{\delta}

\begin{document}
\thispagestyle{empty}
\begin{flushright}
Preprint DFPD 96/TH/45\\
September 1996, hep-th/9609107\\
\end{flushright}

\renewcommand{\thefootnote}{\fnsymbol{footnote}}
\begin{center}
{\large \bf On the BRST Quantization of the Massless Bosonic Particle  \\
in Twistor--Like Formulation.}

\vspace{.3cm}
{\bf Igor Bandos,}

\vspace{0.2cm}
{\it Kharkov Institute of Physics and Technology,}
{\it 310108, Kharkov,  Ukraine}\\
e-mail:  kfti@kfti.kharkov.ua

\vspace{.3cm}
{\bf Alexey Maznytsia, Igor Rudychev}

\vspace{0.2cm}
{\it Department of Physics and Technology,}
{\it Kharkov State University,}
{\it 310108, Kharkov,  Ukraine}

\vspace{0.3cm}
{\bf and}

\vspace{0.3cm}
\renewcommand{\thefootnote}{\dagger}
{\bf Dmitri Sorokin\footnote{on leave from Kharkov Institute of
Physics and Technology, Kharkov, 310108, Ukraine.}},

\vspace{0.2cm}
{\it Universit\`a Degli Studi Di Padova
Dipartimento Di Fisica ``Galileo Galilei''\\
ed INFN, Sezione Di Padova
Via F. Marzolo, 8, 35131 Padova, Italia\\
e--mail: sorokin@pd.infn.it}

\end{center}
\vspace{0.5cm}
\renewcommand{\thefootnote}{\alph{footnote}}
\setcounter{footnote}0
\centerline{\bf Abstract}

\medskip
\noindent
We study some features of bosonic particle
path--integral quantization in a twistor--like approach
by use of the BRST--BFV quantization prescription.
In the course of the Hamiltonian analysis we observe links between
various
formulations of the twistor--like particle
by performing a
conversion of the Hamiltonian constraints of one formulation to another.
A particular feature of the conversion procedure
applied to turn the
second--class constraints into the first--class constraints is that the
simplest Lorentz--covariant way to do this is to convert a
full mixed set of the initial first-- and second--class constraints rather
than explicitly extracting and
converting only the second--class constraints.
Another novel feature of the conversion procedure applied below
is that in the case of the $D=4$ and $D=6$ twistor--like
particle the number of new auxiliary Lorentz--covariant  coordinates,
which one introduces to get a system of first--class constraints in an
extended phase space, exceeds the number
of independent second--class constraints of the original dynamical system.

We calculate the twistor--like particle propagator
in $D=3,4$ and $6$ space--time dimensions
and show, that it coincides with that of a conventional massless bosonic
particle.

\newpage
\setcounter{page}1
\section{Introduction}
In the last decades there has been an intensive activity in studying
(super)particles and (super) strings by use of
different approaches aimed at finding a formulation, which would be the most
appropriate for performing the covariant quantization of the models. Almost
all of the approaches use twistor variables in one form or another
\cite{penrose} -- \cite{bpstv}.
This allowed one to better understand the geometrical and group--theoretical
structure of the theory and to carry out a covariant
Hamiltonian analysis (and in some cases even the covariant quantization)
of (super)particle and (super)string dynamics in space--time
dimensions $D=3,4,6$ and 10, where conventional twistor relations take place.

It has been shown that twistor--like variables appear in a
natural way as superpartners of Grassmann spinor coordinates in a doubly
supersymmetric formulation \cite{spinsup} of Casalbuoni--Brink--Schwarz
superparticles and Green--Schwarz superstrings \cite{gsw}, the notorious
fermionic $\kappa$--symmetry \cite{ks}
of these models being replaced by
more fundamental local supersymmetry on the worldsheet  supersurface  swept
by the superparticles and superstrings in target superspace \cite{stvz}.
This has solved the problem of infinite reducibility of the fermionic
constraints associated with $\kappa$--symmetry \footnote{A
comprehensive
list of references on the subject the reader may find in \cite{bpstv}}.
As a result new formulation and methods of quantization of $D=4$
compactifications of
superstrings with manifest target--space supersymmetry have been
developed (see \cite{ber} for a review).
However, the complete
and simple solution of the problem of $SO(1,D-1)$ covariant quantization
of twistor--like superparticles and superstrings in $D~>~4$ is still lacking.

To advance in solving this problem one has to learn more on how to deal
with twistor--like variables
when performing the Hamiltonian analysis and the quantization of the
models. In this respect a bosonic relativistic particle in a twistor--like
formulation may serve as the simplest but rather nontrivial toy model.

The covariant quantization of the bosonic particle has been under intensive
study with both the operator and path--integral method
\cite{ferber,teit,polyakov,govaerts,sf,sg,bh,bfortschr}.
In the twistor--like approach the bosonic particle has been mainly quantized
by use of the operator formalism. For that different but classically
equivalent twistor--like particle actions have been considered
\cite{ferber,es,sg,bh,bfortschr}.

The aim of the present paper is to study some features of bosonic particle
path--integral quantization in the twistor--like approach
by use of the BRST--BFV quantization prescription \cite{bf} --
\cite{bff}.
In the course of the Hamiltonian analysis we shall observe links between
various
formulations of the twistor--like particle \cite{ferber,es,stvz}
by performing a
conversion of the Hamiltonian constraints of one formulation to another.
A particular feature of the conversion procedure \cite{fs}
applied to turn the
second--class constraints into the first--class constraints is that the
simplest Lorentz--covariant way to do this is to convert a
full mixed set of the initial first-- and second--class constraints rather
than explicitly extracting and converting only the second--class constraints.
Another novel feature of the conversion procedure applied below
(in comparison with the conventional
one \cite{bff,fs}) is that in the case of the $D=4$ and $D=6$ twistor--like
particle the number of new auxiliary Lorentz--covariant  coordinates,
which one introduces to get a system of first--class constraints in an
extended phase space, exceeds the number
of independent second--class constraints of the original dynamical system, 
(but because of an appropriate amount of the first--class constraints
we finally get, the number of physical degrees of freedom remains the 
same).  

In Section 2 the classical mechanics of a twistor--like bosonic 
particle in D=3,4 and 6 is considered.  The Hamiltonian analysis of the 
constraints accompanied by the conversion procedure is carried out and a 
classical BRST charge is constructed by introducing ghosts corresponding 
to a set of the first--class constraints obtained as a result of 
conversion.

        In Section 3 the problem of admissible gauge choice for variables
describing the matter--ghost system of the model is discussed.

        In Section 4 we perform the path--integral quantization of the model
in $D=3,4$ and $6$ space--time dimensions using
the extended BRST scheme \cite{mhenn}. We calculate the particle propagator
and show, that it coincides with that of the massless bosonic particle.
At the end of this Section we make a comment on problems of the D=10
case.

{\it Notation.} We use the following signature for the space-time metrics:
$(+,-,...,-)$.

\section{Classical Hamiltonian dynamics and the BRST-charge.}
\subsection{Preliminaries}
    The  dynamics of a massless bosonic particle in D=3,4,6 and 10 space--time
can be described by the action \cite{ferber}
\be \label{201}
S={1\over 2}\int d\tau \dot x^m ({\bar \l}\gamma _m \l ), \qquad \ee
where $x^m(\tau )$ is a particle space--time coordinate,
$\l ^{\a}(\tau )$ is an
auxiliary bosonic spinor variable, the dot stands for the time derivative
$\partial\over{\partial\tau}$ and  $\gamma ^m $ are the Dirac matrices.

The derivation of the canonical momenta \footnote{In what 
follows $P^{(..)}$ denotes the momentum conjugate to the
variable in  the brackets}
$P^{(x)}_m={{\partial L}
	\over{\partial{\dot x}^m}},~~
P^{(\l )}_{\a}={{\partial L}\over
{\partial}{{\dot \l}^{\a}}}$
results in a set of primary constraints $$
\Psi _m=P^{(x)}_m-{1\over 2}({\bar \l}\gamma _m \l )\approx 0, $$
\be \label{202}
P^{(\l )}_{\a}\approx 0.
\qquad \ee
They form the following
algebra with respect to the  Poisson brackets\footnote{The canonical
Poisson brackets are
$$
[P^{(x)}_m, x^n]_P=\delta ^n_m;
\qquad
[P^{(\l )}_{\a}, \l^{\b}]_P=\delta ^{\b}_{\a} $$}

\be \label{203}
[\Psi _m, \Psi _n]_P=0,~~ [P^{(\l )}_{\a},
P^{(\l )}_{\b}]_P=0,~~ [\Psi _m, P^{(\l )}_{\a}]_P=({\gamma _m}\l )_{\a}.
\qquad \ee
One can check that new independent secondary constraints do not appear
in the model.
In general,
Eqs. \p{202} are a mixture of first-- and  second--class constraints.
The  operator quantization of this
dynamical system in $D=4$ (considered previously in \cite{sg,bh}) was
based on the Lorentz--covariant
splitting of the first-- and second--class constraints and on the subsequent
reduction of the phase space (either by explicit solution of the
second--class constraints \cite{bh} or,
implicitly, by use of the Dirac brackets
\cite{sg}), while
in \cite{bz0,bfortschr} a conversion prescription \cite{bff,fs} was used.
The latter consists in the extension of the phase space of the particle
coordinates and momenta with auxiliary variables in such a way, that new
first--class constraints replace the original second--class ones. Then the
initial
system with the second--class constraints is treated as a gauge fixing of a
 ``virtual" \cite{bff} gauge symmetry generated by the additional first--class
constraints of the extended system \cite{bff,fs}.
This is achieved by taking the auxiliary conversion degrees of freedom to be
zero or expressed in terms of initial variables of the model.

     The direct application of this procedure can encounter some technical
problems for systems, where the first-- and second--class constraints form a
complicated algebra
(see, for example, constraints of the $D=10$ superstring in a
Lorentz--harmonic formulation \cite{bzstr}).
Moreover, in order to perform the covariant
separation of the first-- and second--class constraints in the system
under consideration it is necessary either to introduce one more independent
auxiliary
bosonic spinor $\mu _{\a}$ (the second component of a twistor
$Z^A=(\l^{\a},\mu _{\a})$ \cite{penrose}) or to construct the second twistor
component from the variables at hand by use of a Penrose relation
\cite{penrose} ${\bar \mu}^{\dot \a}=ix^{\a {\dot \a}}\l _{\a} ~(D=4),~
\mu^{\a}=x^{\a \b}\l _{\b} ~(D=3)$. In the latter case the structure of the
algebra of the
first-- and second--class constraints separated this way
\cite{sfortschr,sg} makes the conversion procedure
rather cumbersome. To elude this one can try to simplify the procedure by
converting into the first class the whole set \p{202} of the mixed
constraints. The analogous trick was used to convert fermionic constraints
in superparticle models \cite{es,moshe}.

Upon carrying out the conversion procedure  we get a system characterized by
the set of first--class constraints $T_i$ that form (at least on the mass
shell) a closed algebra  with respect to the Poisson brackets
defined for all the variables of
the modified phase space. In order to perform the BRST--BFV quantization
procedure we associate with each constraint of Grassmann parity $\epsilon $
the pair of canonical conjugate auxiliary variables (ghosts) $\eta _i,~
P^{(\eta )}_i$ with  Grassmann parity $\epsilon +1$ \footnote{If the
extended BRST--BFV method is used, with each constraint  associated are also
a Lagrange multiplier, its conjugate momentum of Grassmann parity $\epsilon
$ and an antighost and its momentum of Grassmann parity $\epsilon +1$ (see
\cite{bf,mhenn} for details).}. The resulting system  is required to be
invariant under gauge transformations generated by a nilpotent fermionic
BRST charge $\Omega $. This invariance substitutes the gauge symmetry,
generated by the first class constraints in the initial phase space.
The generator $\Omega $ is found as a series in
powers of ghosts
$$
\Omega ={\eta _i}T_i + higher~ order~ terms, $$
where the structure of higher--order terms reflects the noncommutative
algebraic structure of the constraint algebra \cite{mhenn}.
Being the generator of the BRST symmetry $\Omega$ must be a dynamical invariant:
$$
{\dot \Omega}=[\Omega ,H]_P=0,
$$
where $H$ is a total Hamiltonian of the system, which has the form
\begin{equation}\label{h}
H=H_0+[\chi ,\Omega]_P.
\end{equation}
In \p{h} $H_0$ is the initial Hamiltonian of the model and $\chi$ is a
gauge fixing fermionic function whose form is determined by admissible gauge
choices \cite{teit,govaerts,sf,west,bfortschr}
(see Section 3 for the discussion of this point).

Upon quantization $\Omega$ and $H$ become operators acting on quantum state
vectors. The physical sector of the model is singled out by the requirement
that the physical states are BRST invariant and vanish under the action of
$\Omega$. Another words, we
deal with a {\it quantum gauge theory}.

When the gauge is fixed, we remain only with physically nonequivalent
states, and the Hamiltonian $H$ is argued to reproduce the correct physical
spectrum of the quantum theory.

        When the model is quantized by the path--integral method, we also
deal with a quantum gauge theory.
The
Hamiltonian \p{h} is used to construct an effective action and a corresponding
BRST-invariant generating functional
which allows one to get transition amplitudes between physical states of the
theory.

        Below we consider the conversion procedure and construct the BRST
charge for the twistor--like particle model in dimensions $D=3,4$ and $6$.

\subsection{ D=3}

In $D=3$ the action \p{201} is rewritten as \begin{equation}\label{211}
  S={1\over 2}\int d\tau \l ^{\a}{\dot x}_{\a \b} \l ^{\b}, \qquad
\end{equation}
where $\l^\a$ is a real two-component commuting spinor (spinor indices are
risen and
lowered by the unit antisymmetric tensor $\epsilon _{\a \b}$) and $x_{\a
\b}=x_m\gamma ^m_{\a \b}$.

The system of
primary constraints \p{202}
\begin{eqnarray} \label{212}
\Psi_{\a \b}=P^{(x)}_{\a \b}-\l_{\a}\l_{\b}\approx 0, \nn  P^{(\l
)}_{\a}\approx 0, \qquad
\end{eqnarray}
is a mixture of a first--class constraint generating the
$\tau$--reparametrization transformations of $x$ $$
\phi = \l^{\a}P^{(x)}_{\a \b}\l^{\b}
$$
and four second--class constraints
\be \label{2125}
(\l P^{(\l )}),~~ (\mu P^{(x)}\mu )-(\l \mu )^{2}, ~~(\mu P^{(\l )}),~~(\l
P^{(x)} \mu ), \qquad \ee where $\mu ^{\a}=x^{\a \b}\l _{\b}$
(see \cite{sg} for details).

In order to perform a conversion of \p{212} into a system of first--class
constraints we introduce a pair of  canonical conjugate bosonic spinors
$(\z^{\a},~P^{(\z )}_{\b})$, $[P^{(\z )}_{\b}, \z^{\a}]_P=\delta ^{\a}
_{\b}$, and take the modified system of constraints, which is of the first
class, in the following form:
\begin{eqnarray} \label{213}
\Psi '_{\a \b}=P^{(x)}_{\a \b}-(\l_{\a}-\z_{\a})(\l_{\b}-\z_{\b})\approx 0,
\nn
\Phi '_{\a}=P^{(\l )}_{\a}+P^{(\z )}_{\a}\approx 0.
\end{eqnarray}

Eqs. \p{213} reduce to \p{212}
by putting the auxiliary variables $\z^{\a}$  and $P^{(\z )}_{\a}$ equal to
zero. This reflects the appearance in the model of a new gauge symmetry with
respect to which $\z^{\a}$  and $P^{(\z )}_{\a}$  are pure gauge degrees of
freedom.

        It is convenient to choose the following phase--space variables as
independent ones:
\begin{eqnarray} \label{214}
v^{\a}=\l^{\a}-\z^{\a}, \qquad
P^{(v)}_{\a}={1\over 2}(P^{(\l )}_{\a}-P^{(\z )} _{\a}), \nn
w^{\a}=\l^{\a}+\z^{\a},
\qquad
P^{(w)}_{\a}={1\over 2}(P^{(\l )}_{\a}+P^{(\z )}_ {\a}), \qquad
\end{eqnarray}
Then Eqs. \p{213} take the following form \begin{eqnarray} \label{215}
\Psi '_{\a \b}=P^{(x)}_{\a \b}-v_{\a}v_{\b}\approx0, \nn P^{(w)}_{\a}\approx
0. \qquad
\end{eqnarray}
These constraints form an Abelian algebra.

One can see that $w^{\a}$ variables do not enter the constraint relations,
and their conjugate momenta are zero. Hence, the quantum physical states of
the model will not depend on $w^{\a}$ .

Enlarging
the modified phase space with ghosts, antighosts and Lagrange multipliers in
accordance with the following table

\begin{tabular}{cccc}
&&&\\
Constraint &  Ghost &     Antighost   &   Lagrange~multiplier   \\
${\Psi}'_{\a \b}$ & $c^{\a \b}$ & $\tilde c^{\a \b}$ & $e^{\a \b}$ \\
$P^{(w)}_{\a}$   &  $b^{\a}$  &   $\tilde b^{\a}$   &  $f^{\a}$ \\
&&&
\end{tabular}

\noindent
we write the classical BRST charges \cite{bf,mhenn} of the model in the
minimal and extended BRST--BFV version as follows \begin{equation} \label{218}
\Omega_{min}=c^{\a \b}\Psi '_{\b \a}+b^{\a}P^{(w)}_{\a}, \qquad \end{equation}
\begin{equation} \label{217}
\Omega =P^{(\tilde c)}_{\a \b}P^{(e)\b \a}+P^{(\tilde b) \a}P^{(f)}_{\a}+
\Omega _{min}. \qquad
\end{equation}

\subsection{D=4}

In this dimension we use two--component $SL(2,C)$ spinors
$(\l^{\a}=\epsilon^{\a \b}\l_{\b}; ~{\bar \l}^{\dot
\a}=\epsilon^{{\dot \a} {\dot \b}}{\bar \l}_{\dot \b}; ~\a ,{\dot
\a}=1,2; ~ \epsilon^{12}=-\epsilon_{21}=1)$. Other notation
coincides with that of the $D=3$ case. Then in $D=4$  the action
\p{201} can be written as following
\begin{equation}\label{221}
S= {1\over 2}\int d \tau \l^\a {\dot x}_{\a {\dot
\a}}\bar{\l^{\dot \a}}, \qquad
\end{equation}
where $ x_{\a\dot \a}= x_{m}{\sigma}^{m}_{\a {\dot \a}} $, and
${\sigma}^m_{\a {\dot
\a}}$ are the relativistic Pauli matrices.  The set of the primary 
constraints \p{202} in this dimension 
$$\Psi_{\a {\dot {\a}}}= P^{(x)}_{\a 
{\dot {\a}}}- \bar{ \l}_{\dot \a} \l_{\a} \approx 0,$$ 
\be \label{222} {  
P^{(\l )}}_{\a}\approx 0, \qquad 
\ee 
$${\bar P}^{(\bar {\l} )}_{\dot 
{\a}}\approx 0 $$ contains two first--class constraints and three pairs of 
conjugate second--class constraints \cite{sg,sfortschr}.  One of the first 
class constraints  generates the $\tau$-reparametrization transformations 
of $x^{{\dot\a} \a}$ $$ \phi= \l^{\a}P^{(x)}_{\a {\dot {\a}}}{\bar 
{\l}}^{\dot {\a}} $$ and another one generates  $U(1)$ rotations of the 
complex spinor variables \be \label{2221} U=i(\l^{\a} P^{(\l )}_\a- \bar 
{\l}^{\dot{\a}}\bar P^{(\l )}_{\dot {\a}}).  \ee

The form of the second--class constraints is analogous to that in the D=3
case (see Eq. \p{2125} and \cite{sg}), and we do not present it explicitly
since it is not used below.

To convert the mixed system of the constraints \p{222} into  first--class
constraints
one should introduce at least three pairs of canonical conjugate auxiliary
bosonic
variables, their number is to be equal to the  number of the second--class
constraints in \p{222}. However, since we do not want to violate the
manifest Lorentz invariance, and the $D=4$ Lorentz group does not have
three--dimensional representations, we are to find a way round.
We introduce two
pairs of canonical conjugate conversion spinors $
(\zeta^{\a},P^{(\zeta )}_{\a}),~~
[\zeta^{\a},P^{(\zeta)}_{\b}]_P=-\delta^{\a}_{\b}, ~~[\bar {\zeta}^{\dot
{\a}},{\bar P}^{(\bar \zeta )}_{\dot {\b}}]_P=-\delta^ {\dot \a}_{\dot \b},
$
(i.e. four pairs of real auxiliary variables) and modify the constraints
\p{222} and the $U(1)$ generator, which becomes an independent first--class
constraint in the enlarged phase space.
Thus we get the following system of the first--class constraints:
$$\Psi^{'}_{\a \dot {\a}}=P^{(x)}_{\a \dot {\a}} -(\bar {\l}-\bar
{\zeta})_{\dot {\a}} (\l-\zeta)_{\a}\approx 0,$$
\be \label{223}
{  \Phi_\a}=P^{(\l )}_\a+P^{(\zeta )}_{\a}\approx 0 ,
\qquad \ee
$$
\bar {\Phi}_{\dot {\a}}= \bar P^{(\bar \l )}_{\dot {\a}}+\bar P^{(\bar \zeta
)}_{\dot {\a}}\approx 0 ,
$$
$$
U=i(\l^{\a} P^{(\l )}_\a+\zeta^{\a}
P^{(\zeta )}_{\a}- \bar {\l}^{\dot {\a}}\bar P^{(\l )}_{\dot {\a}}-\bar
{\zeta}^{\dot {\a}}\bar P^{(\zeta )}_{\dot {\a}})\approx 0.
$$
One can see
(by direct counting), that the number of independent physical
degrees of freedom of the particle  in the enlarged phase space is the
same as in the initial one. The latter is recovered by imposing gauge
fixing conditions on the new auxiliary variables \be \label{2231}
\z^{\a}=0,\qquad{\bar \z}^{\dot \a}=0,\qquad P^{(\z )}_{\a}=0, \qquad
P^{({\bar \z})}_{\dot \a}=0.
\ee

By introducing a new set of the independent spinor variables analogous to
that in \p{214} one rewrites Eqs. \p{223} as follows
$$
\Psi'_{\a {\dot{\a}}}=P^{(x)}_{\a {\dot{\a}}}- v_{\a}{\bar
v}_{\dot{\a}}\approx 0, $$ $$ U=i(P^{(v)}_{\a}v^{\a}- P^{(\bar v)}_{\dot
{\a}}{\bar v}^{\dot {\a}})\approx 0,
$$
\be \label{224}
{  P^{(w)}}_{\a}\approx 0, \qquad \ee $$ P^{(\bar{w})}_{\dot{\a}}\approx 0.
$$
Again, as in the $D=3$ case, $w_\a,~{\bar w}_{\dot\a}$ and their momenta
decouple from the first pair of the constraints \p{224}, and can be
completely excluded from the number of the dynamical degrees of freedom  by
putting
\begin{equation}\label{es}
w_\a=\l_\a+\z_\a=0, \qquad
{P^{(w)}}_{\a}={1\over 2}({ P^{(\l )}}_{\a}+{P^{(\z )}}_{\a})=0 \end{equation}
in the strong sense. This gauge choice, which differs from \p{2231},
reduces
the phase space of the model to that of a version of the twistor--like
particle dynamics, subject to the first pair of the first--class constraints
in \p{224}, considered by Eisenberg and Solomon \cite{es}.
The constraints \p{224} form an abelian algebra, as in the $D=3$ case.
In compliance with the BRST--BFV prescription we introduce ghosts,
antighosts and Lagrange
multipliers associated with the
constraints \p{224} as follows

\begin{tabular}{cccc}
Constraint  &  Ghost &   Antighost &  Lagrange~ multiplier  \\ ${\Psi}
'_{\a {\dot{\a}}}$ &  $c^{\dot{\a} \a}$ & ${\tilde c}^{{\dot \a} \a}$ &
$e^{\dot{\a} \a} $ \\ $U$ & $a$ & ${\tilde a}$ & $g$ \\ $P^{(w)}_{\a}$ &
$b^{\a}$ & ${\tilde b}^{\a}$ & $f^{\a}$ \\ $P^{(\bar{w})}_{\dot{\a}}$ &
${\bar b}^{\dot{\a}}$ & ${\tilde{\bar b}}^{\dot{\a}}$ & ${\bar f}^{\dot
{\a}}$ \\ \qquad \end{tabular}

  Then the BRST--charges of the $D=4$ model have the form
\begin{equation}
\label{227}
\Omega_{min}=c^{\dot {\a} \a}\Psi_{\a {\dot {\a}}}+b^{\a}P^{(w)}_{\a}+
{\bar b}^{\dot{\a}}P^{({\bar w})}_{\dot{\a}}+aU,
\qquad \end{equation}
\be \label{226}
\Omega=P^{({\tilde c})}_{\a {\dot{\a}}}P^{(e)
{\dot{\a}}\a}+P^{({\tilde b})} _{\a}P^{(f)\a}+P^{({\tilde{\bar
b}})}_{\dot{\a}}P^{({\bar f})\dot{\a}}+ P^{({\tilde
a})}P^{(g)}+\Omega_{min}.
\ee

\subsection{D=6}

In $D=6$ a light--like vector $V^m$ can be represented in terms of commuting
spinors as follows
$$
V^m=\l^{\a}_i\gamma^m_{\a \b}\l^{\b i}, $$
where $\l^{\a}_i$ is an $SU(2)$--Majorana--Weyl
spinor which has the $SU^*(4)$ index
$\a =1,2,3,4$ and the $SU(2)$ index $i=1,2$.
$\gamma^m_{\a \b}$ are $D=6$ analogs of the Pauli matrices (see
\cite{6spin,benght}). $SU(2)$ indices are risen and lowered by the unit
antisymmetric tensors $\epsilon_{ij},~~\epsilon^{ij}$. As to the $SU^*(4)$
indices, they can be risen and lowered only in pairs by the totally
antisymmetric tensors $\epsilon_{\a \b \g \d}$, $\epsilon^{\a \b \g \d}$
($\epsilon_{1234}=1)$.

Rewriting the action \p{201} in terms of $SU(2)$--Majorana--Weyl spinors,
one gets
\be \label{231}
S={1\over 2}\int d \tau {\dot x}^m \l^{\a}_i(\gamma_m)_{\a \b}\l^{\b i}.
\ee

   The system of the primary constraints \p{202} takes the form
$$
\Psi_{\a \b}=P^{(x)}_{\a \b}-\epsilon_{\a \b \g \d}\l^{\g}_i \l^{\d i}
\approx 0,
$$
\be \label{232}
{  ~P^{(\l )i}}_{\a}\approx 0, \qquad
\ee
where $P^{(x)}_{\a \b}=P^{(x)}_m \gamma^m_{\a \b}$. $\Psi_{\a \b}$ is
antisymmetric in $\a$ and $\b$ and contains six independent components.
(To get \p{232} we used the relation
$(\g_m)_{\a\b}\g^m_{\g\d}\sim \epsilon_{\a \b \g \d}$).

From Eqs. \p{232}
one can separate four first--class constraints  by projecting  \p{232}
onto $\l^{\a}_{i}$ \cite{sfortschr,benght}.  One of the first--class
constraints generates the $\tau$--reparametrizations of $x^{\a\b}$ $$ \phi
=\l^{\a}_{i}P^{(x)}_{\a \b}\l^{\b i}, $$ and another three ones form an
$SU(2)$ algebra $$ T_{ij}=\l^{\a}_{(i}P^{(\l )}_{\a j)}, $$ Braces denote
the symmetrization of $i$ and $j$.  All other constraints in \p{232} are
of the second class.

   The conversion of \p{232} into first--class constraints is carried out by
analogy  with the $D=4$
case. According to the conventional conversion prescription we had to
introduce five pairs
of canonical conjugate bosonic variables. Instead, in order to preserve
Lorentz invariance, we introduce the canonical conjugate pair of bosonic
spinors $\z^{\b}_j$, $P^{(\z )i}_{\a}$
($[P^{(\z )i}_{\a},\z^{\b}_j]_P=\d^{\b}_{\a}\d^i_j,$) modify the constraints
\p{232} and the $SU(2)$ generators.
This results in the set of independent first--class constraints
$$
\Psi '_{\a \b}=P^{(x)}_{\a \b}-\epsilon_{\a \b \g \d}
(\l^{\g}_i-\z^{\g}_i)(\l^{\d i}-\z^{\d i})\approx 0, $$
\be \label{233}
{  \Phi_{\a}}^i=P^{(\l )i}_{\a}+P^{(\z )i}_{\a}\approx 0, \qquad \ee
$$
T_{ij}=\l^{\a}_{(i}P^{(\l )}_{\a j)}-\z^{\a}_{(i}P^{(\z )}_{\a j)}\approx 0.
$$
In terms of spinors $v^{\a}_{i}$ and $w^{\a}_{i}$, and their momenta,
defined as in the $D=3$ case \p{214}, they take the following form
$$
\Psi '_{\a \b}=P^{(x)}_{\a \b}-\epsilon_{\a \b \g \d}v^{\g}_iv^{\d i}\approx
0, $$
\be \label{234}
{  T_{ij}}=v^{\a}_{(i}P^{(v)}_{\a j)}\approx 0, \qquad \ee
$$ P^{(w)i}_{\a}\approx 0. $$

    These constraints form a closed algebra with respect to the Poisson
brackets.
The only nontrivial brackets in this algebra are \be \label{235}
[T_{ij},T_{kl}]_{p}=\epsilon_{jk}T_{il}+\epsilon_{il}T_{jk}+
\epsilon_{ik}T_{jl} +\epsilon_{jl}T_{ik}, \qquad \ee
which generate the $SU(2)$ algebra.

We introduce ghosts, antighosts and Lagrange multipliers related to the
constraints \p{235}

\begin{tabular}{cccc}
& & &\\
Constraint  &  Ghost  &  Antighost &  Lagrange~ multiplier  \\ ${\Psi} '_{\a
\b}$ & $c^{\a \b}$ & ${\tilde c}_{\a \b}$ & $e^{\a \b}$  \\ $T_{ij}$ &
$a^{ij}$ & ${\tilde a}_{ij}$ & $g^{ij}$ \\ $\Phi ^{i}_{\a}$ & $b^{\a}_{i}$ 
& ${\tilde b}^{i}_{\a}$ & $f^{\a}_{i}$ \\ &&& \end{tabular}

\noindent
and construct the BRST charges corresponding respectively, to the minimal
and extended BRST--BFV version, as follows \begin{equation}
\label{238}
\Omega_{min}=c^{\a \b}\Psi '_{\b \a}+b^{\a}_{i}P^{(w)i}_{\a}+a^{ij}T_{ji}+
\qquad
\ee
$$
(\epsilon_{jk}P^{(a)}_{il}+ \epsilon_{il}P^{(a)}_{jk}+
\epsilon_{ik}P^{(a)}_{jl}+\epsilon_{jl}P^{(a)}_{ik})a^{ij}a^{kl}.
$$
\be \label{237}
\Omega=P^{({\tilde c})}_{\a \b}P^{(e)\b \a}+P^{({\tilde b})\a}
_{i}P^{(f)i}_{\a}+P^{({\tilde a})ij}P^{(g)}_{ji}+\Omega_{min}, \qquad \ee
Higher order terms in ghost powers appear in \p{238} and \p{237} owing to the
noncommutative $SU(2)$ algebra of the $T_{ij}$ constraints \p{235}.

\section{Admissible gauge choice.}

   One of the important problems in the quantization of gauge
systems is a correct gauge choice. In the frame of the BRST--BFV
quantization scheme gauge fixing is made by an appropriate choice of the
gauge fermion that determines the structure of the quantum Hamiltonian.
The Batalin and Vilkovisky theorem \cite{bf,mhenn} reads that the result of
path integration does not depend on the choice of the gauge fermions if they
belong
to the same equivalence class with respect to the BRST--transformations.
An analogous theorem
takes place in the operator BRST--BFV quantization scheme \cite{sf}.
Further
analysis of this problem for systems possessing the reparametrization
invariance showed that the result of path integration does not depend on
the choice of the
gauge fermion if only appropriate
gauge conditions are compatible with the  boundary conditions for the
parameters of the corresponding gauge transformations
\cite{polyakov,govaerts,sf,west,bfortschr}. In particular, it was shown
that the so--called ``canonical gauge", when the worldline gauge field of the
reparametrization symmetry of the bosonic particle is fixed to be a
constant, is not admissible in this sense.
(see \cite{govaerts,bfortschr}
for details). Anyway one can use the canonical gauge as a consistent limit
of an admissible gauge \cite{sf}.

Making the analysis of the twistor--like model one can show that admissible
are the following gauge conditions on Lagrange multipliers from the
corresponding Tables of the previous section in the dimensions $D=3,$ 4 and
$6$ of space--time, respectively,
\be \label{261}
D=3:\qquad {\dot e}^{\a\b}=0;\qquad f^{\a}=0; \ee
\be \label{262}
D=4:\qquad {\dot e}^{\a\dot\b}=0;\qquad f^{\a}=0; \qquad f^{\dot
\a}=0;\qquad g=0;
\ee
\be \label{263}
D=6:\qquad {\dot e}^{\a\b}=0;\qquad f_i^{\a}=0;\qquad g^{ij}=0; \ee

The canonical gauge
\be \label{264}
e=constant,
\ee
can be considered as a limit of
more general admissible gauge $e-\varepsilon {\dot e}=constant$ (at
$\varepsilon \rightarrow 0$) \cite{sf}.
Then the use of the gauge condition \p{264} does not lead to any problems
with the operator BRST--BFV quantization.

Below we shall use the ``relativistic" gauge conditions \p{261}, \p{262} and
\p{263} for the path--integral quantization. The use of the canonical
gauge \p{264} in this case would lead to a wrong form of the particle
propagator.

\section{Path--integral BRST quantization.}

In this section we shall use the
extended version of the BRST--BFV quantization procedure \cite{mhenn,bff}
and fix the gauge by applying the conditions \p{261}, \p{262}, \p{263}.
The gauge fermion, corresponding to this gauge choice, is
     \be \label{417}
     \chi_D={1\over 2}P^{(c)}_me^m, ~~~D=3,4,6, \qquad
     \ee

The Hamiltonians constructed with \p{417} are \cite{bf,mhenn}
$$ {\it H}_D=[\Omega_D,\chi_D],\qquad D=3,4,6 $$ \be \label{420}
{H}_3=e^{m}(P^{(x)}_{m}-{1\over 2}v^{\a}(\gamma_{m})_{\a \b}v^{\b})
-P^{(c)}_{m}P^{({\tilde c})m}, \qquad
\ee
\be \label{421}
{H}_4=e^{m}(P^{(x)}_{m}-{1\over 2}{\bar v}^{\dot {\a}}(\sigma_{m})
_{\dot{\a}\a}v^{\a})-P^{(c)}_{m}P^{({\tilde c})m}, \qquad \ee
\be \label{422}
{H}_6=e^{m}(P^{(x)}_{m}-{1\over 2}v^{\a}_i(\g_{m})_ {\a \b}v^{\b i})
-P^{(c)}_{m}P^{({\tilde c})m}, \qquad \ee

We shall calculate the coordinate propagator
$Z=\langle x_{1}^{m}\mid U_0\mid x_{2}^{m}\rangle$ (where 
$U_0=expiH(T_1-T_2)$ is the evolution operator), 
therefore boundary conditions for the phase space 
variables are fixed as follows:  \be \label{423} x^{m}(T_1)=x_{1}^{m}, 
\qquad x^{m}(T_2)=x_{2}^{m}, \qquad \ee the boundary values of the ghosts, 
antighosts and canonical momenta of the Lagrange multipliers are put equal 
to zero (which is required by the BRST invariance of the boundary 
conditions \cite{mhenn}), and we sum up over all possible values of the 
particle momentum and the twistor variables.

The standard expression for the matrix element of the evolution  operator is
\be \label{424}
 Z_D=\int[D\mu DP^{\mu}]_D
 exp(i\int_{T_1}^{T_2} d \tau ([P^{\mu}{\dot {\mu}}]_D -{\it H}_D )),
\qquad D=3,4,6.
\ee
$[D\mu DP^{\mu}]_D $ contains functional Liouville measures
of all the canonical variables of the BFV extended phase space \cite{bf}.
$[P^{\mu}{\dot{\mu}}]_D $ contains a sum of products
of the canonical momenta with the velocities.

For instance, an explicit expression for the path--integral measure
in the $D=3$ case is
$$
[D\mu DP^{\mu}]=DxDP^{(x)}DvDP^{(v)}DwDP^{(w)}DeDP^{(e)}DfDP^{(f)} $$
$$
DbDP^{(b)}DcDP^{(c)}D{\tilde b}DP^{({\tilde b})}D{\tilde c}DP^{({\tilde
c})}.
$$

We can perform straightforward integration over  the all variables
that are not  present in the Hamiltonians \p{420}, \p{421}, \p{422}
\footnote{ All calculations are done up to a multiplication constant,
which can always be absorbed by the integration measure.}.
Then \p{424} reduces to the product of two terms
\be \label{425}
 Z_D=I_DG_D, \qquad
\ee
where
\be \label{426}
G_D=\int DcDP^{(c)}D{\tilde c}DP^{({\tilde c})} exp(i\int_{T_1}^{T_2}
d \tau (P^{(c)}_m{\dot c}^m+P^{({\tilde c})}_m{\dot{\tilde c}}^m
-{1\over 2} P^{({\tilde c})}_mP^{(c)m})), \qquad
\ee
and $I_D$ includes the integrals over bosonic variables entering \p{420},
\p{421}, \p{422} together with their conjugated momenta.
We use the method analogous to that in \cite{rivelles}
for computing these integrals.

The calculation of the ghost integral $G_D$ results in \be \label{427}
G_D=(\Delta T)^D, \qquad
\Delta T=T_2-T_1, \qquad D=3,4,6.
\ee
Let us demonstrate main steps of the $I_D$ calculation in the $D=3$
case
\begin{eqnarray}\label{801}
I_3&=&\int DxDP^{(x)}DeDP^{(e)}DvDP^{(v)}exp(i\int _{T_1}^{T_2}d\tau
(P_m^{(x)}{\dot x}^m+P_m^{(e)}{\dot e}^m+
P_{\a}^{(v)}{\dot v}^{\a} \nn
& & -e^{m}(P^{(x)}_{m}-
{1\over 2}v^{\a}(\g _m)_{\a \b} v^{\b}))
\end{eqnarray}
Integration over $P_m^{(e)}$ and $P_m^{(v)}$
results in the functional $\delta$-functions $\delta ({\dot e}),~ \delta
({\dot v})$ which reduce functional integrals over $e^m$ and $v^{\a}$ to
ordinary ones:
\be \label{802}
I_3=\int DxDP^{(x)}d^{3}ed^{2}v~exp(ip_m\Delta x^m- \\
i\int _{T_1}^{T_2}d\tau (x^m{\dot P}^{(x)}_m+e^m(P^{(x)}_m-
{1\over 2}v^{\a}(\g _m)_{\a \b}v^{\b})),
\qquad \ee
where $\Delta x^m=x_2^m-x_1^m$ \p{423}. Since the
integral over $v^{\a}$ is a usual Gauss integral after
integrating over $x^m$ and $v^{\a}$ one obtains
\be \label{803}
I_3=\int d^{3}pd^{3}e{1\over {\sqrt{e^me_m-i0}}}exp(i(p_m\Delta x^m-
e^mp_m\Delta T)).
\ee
In general case of $D=3$, 4 and 6 dimensions, one obtains
\be \label{428}
I_D=\int d^Dpd^De{{1\over{(e^me_m-i0)^{{D-2}\over 2}}}} exp(i(p_m\Delta
x^m- e^mp_m\Delta T)),
\ee
that can be rewritten as
\be \label{804}
I_D=\int
d^{D}pd^{D}e\int_0^{\infty}dc~exp(i(p_m\Delta x^m-e^mp_m\Delta T
+(e^me_m-i0)c^{2\over {D-2}})),
\qquad \ee
where $c$ is an auxiliary variable.

Integrating over  $p^m$ and $e^m$ one gets
\be \label{436}
Z_D=\int_{0}^{\infty}dc{1\over{c^{D/2}}}
exp(i{{\Delta x^m\Delta x_m}\over{2c}} -c0), \qquad D=3,4,6,
\ee

or
$$
Z_D={1\over {(\Delta x^m\Delta x_m-i0)^{{D-2}\over 2}}}, $$
which coincides with  the coordinate propagator
for the massless bosonic particle in the
standard formulation \cite{govaerts}.

On the other hand integrating \p{428} only over $e^m$ we get
the massless bosonic particle causal propagator in the form $$
Z_D=\int d^Dp{1\over{p^mp_m+i0}}exp(ip_m\Delta x^m).
$$

\subsection{Comment on the $D=10$ case}
Above we have restricted our consideration to the space--time dimensions
3, 4 and 6. The case of a bosonic twistor--like particle in $D=10$ is
much more sophisticated. The Cartan--Penrose representation of a
$D=10$ light--like momentum vector is constructed out of a
Majorana--Weyl spinor $\lambda^\a$ which has 16 independent components
\be\label{pc}
P^m=\l\Gamma^m\l.
\ee
Transformations of $\l^\a$ which leave \p{pc} invariant take values on
an $S^7$-- sphere (see \cite{es,nispach,bfortschr}
and references therein). In contrast to
the $D=4$ and $D=6$ case, where such transformations belong to the
group $U(1)\sim S^1$ \p{224} and $SU(2)\sim S^3$ \p{233},
respectively, $S^7$
is not a Lie group and its corresponding algebra contains structure
functions instead of structure constants. Moreover, among the 10
constraints \p{pc}  and 16 constraints $P^{(\l)}_\a=0$ on the
momenta conjugate to $x^m$ and $\l^\a$ $18=10+16-1-7$ (where 7 comes from
$S^7$ and 1 corresponds to local $\tau$--reparametrization) are of the
second class. They do not form a representation of the Lorentz group
and cause the problem for covariant Hamiltonian analysis.

One can overcome these problems in the framework of the
Lorentz--harmonic formalism (see \cite{bzstr,bpstv} and references 
therein), where to construct a light--like vector one introduces eight 
Majorana--Weyl spinors instead of one $\l^\a$. Such a spinor matrix
takes values in a spinor representation of the  double covering group
$Spin(1,9)$ of $SO(1,9)$ and satisfies second--class harmonic
conditions. The algebra of the constraints in this ``multi--twistor"
case is easier to analyze than that with only one commuting spinor
involved. The path--integral
BRST quantization of the $D=10$ twistor--like particle is in progress.

\section {Conclusion}
In the present paper the BRST--BFV quantization
of the dynamics of massless bosonic particle in $D=3,4,6$
was performed in the twistor--like formulation.
To this end the initially mixed system of the first-- and
second--class constraints was converted into the system of
first--class constraints by extending the initial phase space
of the model with auxiliary variables in a Lorentz--covariant way.
The conversion procedure (rather than having been a formal trick)
was shown to have a meaning of a symmetry transformation
which relates different twistor--like formulations of the bosonic particle,
corresponding to different gauge choices in the extended phase space.

We quantized the model by use of
the extended BRST--BFV scheme for the path--integral quantization.
As a result we have
presented one of the numerous proofs of the equivalence between
the twistor--like and conventional formulation of
the bosonic particle mechanics.

This example demonstrates peculiar features of treating the
twistor--like variables within the course of the covariant
Hamiltonian analysis and the BRST quantization,
which one should take into account when studying more
complicated twistor--like systems, such as superparticles and
superstrings.

\bigskip
\noindent
{\bf Acknowledgements.}

\noindent
The authors are grateful to P. Pasti, M. Tonin, D. V. Volkov and V. G. Zima
for useful discussion. I. Bandos and D. Sorokin acknowledge partial
support from the INTAS and Dutch Government 
Grant N 94--2317 and the INTAS Grant N 93--493.

\end{document}